\documentclass[pra,twocolumn,tightenlines,showpacs,nofootinbib]{revtex4}
\usepackage{multirow}
\usepackage{bm,dcolumn,amsmath,graphicx}

\newcommand{\eref}[1]{Eq.~(\ref{#1})}

\begin{document}
\title{Blackbody radiation shift in the Sr optical atomic clock}

\author{M.~S.~Safronova$^1$}
\author{S.~G.~Porsev$^{1,2}$}
\author{U.~I.~Safronova$^{3,4}$}
\author{M. G. Kozlov$^{2,5}$}
\author{Charles W. Clark$^6$}
\affiliation{ $^1$Department of Physics and Astronomy, University of Delaware,
    Newark, Delaware 19716, USA\\
$^2$Petersburg Nuclear Physics Institute, Gatchina, Leningrad District, 188300, Russia \\
$^3$Physics Department, University of Nevada, Reno, Nevada 89557,
\\$^4$Department of Physics,  University of Notre Dame, Notre Dame,
IN 46556 \\
$^5$St. Petersburg Electrotechnical University ``LETI'', Prof. Popov Str. 5, St. Petersburg, 197376, Russia\\
$^6$Joint Quantum Institute, National Institute of Standards
and Technology and the \\University of Maryland, Gaithersburg,
Maryland, 20899-8410, USA}
\date{\today}

\begin{abstract}
We evaluated the static and dynamic polarizabilities of the
$5s^2~^1$S$_0$ and $5s5p~^3$P$_0^o$ states of Sr using the
high-precision relativistic configuration interaction + all-order
method. Our calculation explains the discrepancy between the recent
experimental $5s^2~^1$S$_0 - 5s5p~^3$P$_0^o$ dc Stark shift
measurement $\Delta \alpha = 247.374(7)$ [Middelmann \textit{et. al},
arXiv:1208.2848 (2012)] and the earlier theoretical result of 261(4)
a.u. [Porsev and Derevianko, Phys. Rev. A {\bf 74}, 020502R (2006)].
Our present value of 247.5~a.u. is in excellent agreement with the
experimental result. We also evaluated  the dynamic correction to the
BBR shift with 1\% uncertainty;  -0.1492(16)~Hz. The dynamic
correction to the BBR shift is unusually large in the case of Sr
(7\%) and it enters significantly into the uncertainty budget of the
Sr optical lattice clock.  We suggest future experiments that could
further reduce the present uncertainties.
\end{abstract}

\maketitle

\section{Introduction}
Optical lattice clocks have shown tremendous progress in recent years
\cite{SwaMarBis12}. An optical frequency standard based on the
$5s^2~^1$S$_0 - 5s5p~^3$P$_0^o$ transition of  ultracold  $^{87}$Sr
atoms confined in a one-dimensional optical lattice is pursued by a
number of
groups~\cite{CamLudBla08,LudZelCam08,FalSchWin11,BaiFouLet08,HonMusTak09,CurOvcHil09}.
 Its systematic uncertainty has been demonstrated at the
10$^{-16}$ fractional frequency level  and an order-of-magnitude
improvement is expected to be achieved
soon~\cite{SwaMarBis12,FalSchWin11}. A three-dimensional optical
lattice clock with bosonic $^{88}$Sr was demonstrated for the first
time in ~\cite{AkaTakKat08}.

The measured clock transition frequencies must be corrected in
practice for the effect of the ambient blackbody radiation (BBR)
shift, which is quite difficult to measure directly. The BBR shift
can only be suppressed by cooling the clock. At room temperature, the
differential BBR shift of the two levels of a clock transition turns
out to make one of the largest irreducible contributions to the
uncertainty budget of optical atomic clocks. The Sr clock transition
 has the largest BBR shift of all optical frequency standards that
are currently under development (see Ref.~\cite{SafKozCla12} for a
recent review). The fractional BBR shift ${\Delta\nu_{\rm
BBR}}/{\nu_0}$ in Sr is more than a factor of 1000 larger than the
fractional BBR shift in the Al$^+$ ion clock~\cite{SafKozCla11}. The
BBR shift of an optical clock can generally be approximated by the dc
Stark shift of the clock transition to about 1-2\% precision, because
optical frequencies are ~ 100 times greater than characteristic BBR
frequencies. However, Sr represents an exception, where the so-called
dynamic correction \cite{PorDer06BBR}, that needs to be determined
separately from the dc Stark shift, is 7\%.
\begin{table*}
\caption{Comparison of experimental \cite{RalKraRea11} and theoretical
energy levels of Sr in cm$^{-1}$.
Two-electron binding energies are given in the first row, energies in other rows are given relative to the ground
state. Results of the  CI+MBPT and CI+all-order calculations are given in columns labeled ``CI+MBPT'' and ``CI+All (A)''. The
CI+all-order values with the ground state two-electron binding energy shifted by 200~cm$^{-1}$ are given in column
labeled ``CI+All (B)''. Corresponding relative differences of these three calculations with experiment are given in the
three corresponding columns labeled ``Diff.'' in \%. The $5s4d~^3$D$_1 - 5s5p~^3$P$_0^o$ transition energy is given in
the last row.}
\label{table1}
\begin{ruledtabular}
\begin{tabular}{lccccccc}
\multicolumn{1}{l}{State} & \multicolumn{1}{c}{Expt.} & \multicolumn{1}{c}{CI+MBPT} & \multicolumn{1}{c}{Diff.(\%)} &
\multicolumn{1}{c}{CI+All (A)} & \multicolumn{1}{c}{Diff.(\%)} &
\multicolumn{1}{c}{CI+All (B)} & \multicolumn{1}{c}{Diff.(\%)} \\
\hline
      $5s^2~^1$S$_0$& 134897 & 136244 & 1.00 &   135444 & 0.41 &   135244&  0.26  \\
      $5s4d~^3$D$_1$& 18159  & 18225  & 0.36 &   18327  & 0.93 &   18127 &  $-$0.18  \\
      $5s4d~^3$D$_2$& 18219  & 18298  & 0.44 &   18394  & 0.96 &   18194 &  $-$0.13  \\
      $5s4d~^3$D$_3$& 18319  & 18422  & 0.56 &   18506  & 1.02 &   18306 &  $-$0.07  \\
      $5s4d~^1$D$_2$& 20150  & 20428  & 1.38 &   20441  & 1.45 &   20241 &  0.45   \\
      $5s6s~^3$S$_1$& 29039  & 29369  & 1.14 &   29223  & 0.63 &   29023 &  $-$0.06  \\
      $5s6s~^1$S$_0$& 30592  & 30938  & 1.13 &   30777  & 0.61 &   30577 &  $-$0.05   \\
      $5s5d~^1$D$_2$& 34727  & 35092  & 1.05 &   34958  & 0.66 &   34758 &  0.09    \\
      $5s5d~^3$D$_1$& 35007  & 35371  & 1.04 &   35210  & 0.58 &   35010 &  0.01    \\
      $5s5d~^3$D$_2$& 35022  & 35388  & 1.04 &   35226  & 0.58 &   35026 &  0.01    \\
      $5s5d~^3$D$_3$& 35045  & 35412  & 1.05 &   35250  & 0.59 &   35050 &  0.01   \\
      $5p^2~^3$P$_0$& 35193  & 35854  & 1.88 &   35545  & 1.00 &   35345 &  0.43    \\
      $5p^2~^3$P$_1$& 35400  & 36070  & 1.89 &   35758  & 1.01 &   35558 &  0.45    \\
      $5p^2~^3$P$_2$& 35675  & 36344  & 1.88 &   36039  &1.02  &   35839 &  0.46   \\
      $5s7s~^3$S$_1$& 37425  & 37776  & 0.94 &   37606  & 0.48 &   37406 &  $-$0.05   \\
      $5s6d~^3$D$_1$& 39686  & 40050  & 0.92 &   39876  & 0.48 &   39676 &  $-$0.02    \\[0.5pc]
      $5s5p~^3$P$_0^o$& 14318&   14806&   3.41  &  14550 &  1.62 &   14350 &  0.23   \\
      $5s5p~^3$P$_1^o$& 14504&   14995&   3.38  &  14739 &  1.61 &   14539 &  0.24    \\
      $5s5p~^3$P$_2^o$& 14899&   15399&   3.36  &  15142 &  1.63 &   14942 &  0.29    \\
      $5s5p~^1$P$_1^o$& 21698&   21955&   1.18  &  21823 &  0.57 &   21623 &  $-$0.35   \\
      $4d5p~^3$F$_2^o$& 33267&   33719&   1.36  &  33648 &  1.14 &   33448 &  0.54    \\
      $4d5p~^3$F$_3^o$& 33590&   34089&   1.49  &  34003 &  1.23 &   33803 &  0.64    \\
      $4d5p~^3$F$_4^o$& 33919&   34444&   1.55  &  34347 &  1.26 &   34147 &  0.67     \\
      $4d5p~^1$D$_2^o$& 33827&   34218&   1.16  &  34208 &  1.13 &   34008 &  0.54     \\
      $5s6p~^3$P$_0^o$& 33853&   34241&   1.15  &  34055 &  0.59 &   33855 &  0.00     \\
      $5s6p~^3$P$_1^o$& 33868&   34255&   1.14  &  34071 &  0.60 &   33871 &  0.01     \\
      $5s6p~^3$P$_2^o$& 33973&   34365&   1.15  &  34134 &  0.47 &   33934 &  $-$0.12   \\
      $5s6p~^1$P$_1^o$& 34098&   34476&   1.11  &  34308 &  0.62 &   34108 &  0.03
      \\[0.5pc]
$^3$D$_1 - ^3$P$_0^o$& 3842 &   3419 &   $-$11.0&  3777   & $-$1.69 &  3777   &  $-$1.69         \\
\end{tabular}
\end{ruledtabular}
\end{table*}
Recently, the dc Stark shift in Sr has been measured with 0.003\% precision ~\cite{MidFalLis12}, and the dynamic
correction was evaluated based on a set of E1 transition rates and the Stark shift measurement. The measured value
differed substantially (by almost $4\sigma$) from the previous theoretical determination ~\cite{PorDer06BBR}.

 In this work, we evaluate the static and dynamic
polarizabilities of the $5s^2~^1$S$_0$ and $5s5p~^3$P$_0^o$ states of
Sr using the high-precision relativistic CI+all-order method. Our
calculation explains the above-mentioned discrepancy between the
 experimental $5s^2~^1$S$_0 - 5s5p~^3$P$_0^o$ dc Stark shift
measurement $\Delta \alpha = 247.374(7)$~a.u.~\cite{MidFalLis12} and
the earlier theoretical result of 261(4) a.u.~\cite{PorDer06BBR}. We
found  that the E1 matrix elements for the transitions that give
dominant contributions to the $^3$P$_0^o$ polarizability, in
particular the $5s4d~^3$D$_1 - 5s5p~^3$P$_0^o$, are rather sensitive
to the higher-order corrections to the wave functions and other
corrections to the matrix elements beyond the random phase
approximation. A correction of only  2.4\%  to the dominant $^3$D$_1
- ^3$P$_0^o$ matrix element leads to 5\% difference in the final
value of the $^3$P$_0^o -\, ^1$S$_0$ Stark shift. In this work, we
included the higher-order corrections in an \textit{ab initio} way
using the CI+all-order approach, and also calculated several other
corrections omitted in~\cite{PorDer06BBR}. Our value for the dc Stark
shift of the clock transition, 247.5~a.u., is in excellent agreement
with the experimental result 247.374(7) a.u.~\cite{MidFalLis12}.

 We have combined
our theoretical calculations with the experimental measurements of
the Stark shift \cite{MidFalLis12} and magic wavelength
\cite{LudZelCam08} of the $5s^2~^1$S$_0 - 5s5p~^3$P$_0^o$ transition
to infer  recommended values of the several electric-dipole matrix
elements that give the dominant contributions to the $^3$P$_0^o$
polarizability. We used these values to evaluate the dynamic
correction to the BBR shift of the $^1$S$_0 -\, ^3$P$_0^o$ transition
to be -0.1492(16)~Hz.

We determined that the $5s4d~^3$D$_1 - 5s5p~^3$P$_0^o$ transition
contributed 98.2\% to the dynamic correction for the $^3$P$_0^o$
level. Our calculation enables us to propose an approach for further
reduction of the uncertainty in the BBR shift. In particular, there
is a correlation in the uncertainty of the BBR shift and the lifetime
of the $5s4d~^3$D$_1$ state, if branching ratios are known to
sufficient accuracy. At present, experimental measurements of the
$5s4d~^3$D term-averaged lifetime have an uncertainty of about 7\%
\cite{MilYuCoo92,RedSanEci04}. We note that the experiment
\cite{MilYuCoo92}, which was performed at JILA some 20 years ago, has
relevance in the determination of the uncertainty budget of one of
the world's most accurate clocks now being developed at the same
institution - a development probably not envisaged at the time.

 A new determination of this (or
$^3$D$_1$) lifetime with 0.5\% uncertainty would provide a value of
the Sr clock BBR shift that is accurate to about 0.5\%, which would
be a factor of 2 improvement in the uncertainty that we state here.
This result is determined by the relevant branching ratios needed for
the extraction of the $5s4d~^3$D$_1 - 5s5p~^3$P$_0^o$ matrix elements
from the
 lifetime measurement.  We have determined these branching ratios
  with an uncertainty of 0.2\%.  A further reduction in the uncertainty
  of the Sr clock
  BBR shift could be effected by an improved measurement of these branching ratios.
The lifetime of the corresponding $6s5d~^3$D$_1$ state in Yb has been
recently measured in Ref.~\cite{Ybtau}.

\begin{table*}
\caption{The CI+MBPT and CI+all-order results and further corrections to the E1 matrix elements for transitions that
give dominant contributions to the polarizabilities of the $5s^2~^1$S$_0$ and $5s5p~^3$P$_0^o$ states. The CI+MBPT and
CI+all-order results including RPA corrections are given in columns labeled ``MBPT+RPA'' and ``All+RPA'', respectively.
The relative differences between the CI+all-order+RPA and CI+MBPT+RPA results are given in column ``Higher orders'' in
\%. The other contributions include  the core-Brueckner ($\sigma$), two-particle (2P), structural radiation (SR), and
normalization (Norm) corrections. Total relative size of corrections beyond CI+all-order+RPA is given in column
``Corr.'' in \%. The recommended values for the  $5s^2~^1$S$_0 - 5s5p~^1$P$_1^o$ matrix element was obtained
from the $^1\!P_1^o$ lifetime measurement~\cite{YasKisTak06}, and the recommended values for all other transitions are from the present work (see Section~\ref{recomm}).}
\label{table2}
\begin{ruledtabular}
\begin{tabular}{lccdddddddc}
\multicolumn{1}{c}{Transition} & \multicolumn{1}{c}{MBPT+RPA} & \multicolumn{1}{c}{All+RPA} &  \multicolumn{1}{c}{Higher orders} &\multicolumn{1}{c}{2P} & \multicolumn{1}{c}{$\sigma$} &  \multicolumn{1}{c}{SR} & \multicolumn{1}{c}{Norm} & \multicolumn{1}{c}{Final} & \multicolumn{1}{c}{Corr.(\%)}& \multicolumn{1}{c}{Recomm.}\\
\hline
 $5s^2~^1$S$_0  - 5s5p~ ^1$P$_1^o$ &5.253 & 5.272 & 0.36\%   &    -0.006 & 0.004  & 0.032    &   -0.094 & 5.208 & -1.23 & 5.248(2)\cite{YasKisTak06}\\
 $5s5p^3$P$_0^o - 5s4d~ ^3$D$_1$   &2.681 & 2.712 & 1.14\%   &    -0.016 & 0.003  & 0.015    &   -0.048 & 2.667 & -1.69 & 2.675(13)\\
 $5s5p^3$P$_0^o - 5s6s~ ^3$S$_1$   &1.983 & 1.970 & -0.66\% &   0.002  & -0.001 & -0.006 &   -0.025 & 1.940 & -1.55 & 1.962(10)\\
 $5s5p^3$P$_0^o - 5s5d~ ^3$D$_1$   &2.474 & 2.460 & -0.57\% &   0.007  & -0.001 & -0.003 &   -0.031 & 2.432 & -1.15 & 2.450(24)\\
 $5s5p^3$P$_0^o - 5p^2~ ^3$P$_1$   &2.587 & 2.619 & 1.22\%    &   0.009  & 0.003    & 0.021    &   -0.033 & 2.620 &    0.04 & 2.605(26)
\end{tabular}
\end{ruledtabular}
\end{table*}

\section{Method and energy levels}

Calculation of  Sr  properties requires an accurate all-order treatment of electron correlations. This can be
accomplished within the framework of the CI+all-order method  that combines configuration interaction and
coupled-cluster approaches \cite{SafKozJoh09,SafKozCla11,PorSafKoz12,SafPorKoz12,SafPorCla12}. To evaluate
uncertainties of the final results, we also carry out CI \cite{KotTup87} and CI+many-body perturbation theory (MBPT)
\cite{DzuFlaKoz96b} calculations.
 These methods have been described in a number
of papers \cite{KotTup87,DzuFlaKoz96b,SafKozJoh09,SafKozCla11} and
 we provide only a brief outline of these approaches.

We start with a solution of the Dirac-Fock  (DF) equations
 $$ H_0\, \psi_c = \varepsilon_c \,\psi_c, $$
  where
$H_0$ is the relativistic DF Hamiltonian \cite{DzuFlaKoz96b,SafKozJoh09}
 and $\psi_c$ and
$\varepsilon_c$ are single-electron wave functions and energies. The calculations are
 carried out in the
$V^{\rm{N-2}}$ potential. The wave functions and the low-lying energy
levels are determined by solving the multiparticle relativistic
equation for two valence electrons~\cite{KotTup87}, $$ H_{\rm
eff}(E_n) \Phi_n = E_n \Phi_n. $$ The effective Hamiltonian is
defined as $$ H_{\rm eff}(E) = H_{\rm FC} + \Sigma(E), $$ where
$H_{\rm FC}$ is the Hamiltonian in the frozen-core approximation. The
energy-dependent operator $\Sigma(E)$ which takes into account
virtual core excitations is constructed using second-order
perturbation theory in the CI+MBPT method \cite{DzuFlaKoz96b} and
using a linearized coupled-cluster single-double method in the
CI+all-order approach \cite{SafKozJoh09}. It is zero in a pure CI
calculation. We refer the reader to
Refs.~\cite{DzuFlaKoz96b,SafKozJoh09} for detailed description of the
construction of the effective Hamiltonian.

 Unless stated otherwise, we use atomic units (a.u.) for all matrix
elements and polarizabilities throughout this paper: the numerical values of the elementary
 charge, $|e|$, the reduced Planck constant, $\hbar = h/2
\pi$, and the electron mass, $m_e$, are set equal to 1. The atomic unit for polarizability can be
converted to SI units via $\alpha/h$~[Hz/(V/m)$^2$]=2.48832$\times10^{-8}\alpha$~(a.u.), where the
conversion coefficient is $4\pi \epsilon_0 a^3_0/h$ and the Planck constant $h$ is factored out in
order to provide direct conversion into frequency units; $a_0$ is the Bohr radius and $\epsilon_0$
is the electric constant.

As a first test of the accuracy of our calculations, we compare our
theoretical energies with experiment for a number of the even and odd
parity states. Comparison of the energy levels (in cm$^{-1}$)
obtained in the CI+MBPT, and CI+all-order approximations with
experimental values \cite{RalKraRea11} is given in
Table~\ref{table1}. Ground state two-electron binding energies are
given in the first row of Table~\ref{table1}, energies in other rows
are measured from the ground state. The relative differences of the
CI+MBPT and CI+all-order
  calculations with experiment (in \%) are given in  columns labeled ``Diff''.
  Since the  CI+all-order
values are systematically  higher than the experimental values, a
large fraction of the difference from experiment can be attributed to
the difference in the value of the ground state two-electron binding
energy. We find that shifting the CI+all-order value of the ground
state two-electron binding energy by only 200~cm$^{-1}$ (see results
in column CI+all-order (B))   brings the results into excellent
agreement with experiment for most of the states. We give the $5s4d
~^3$D$_1 - 5s5p~^3$P$_0^o$ transition energy in the last row of
Table~\ref{table1}. This transition is particulary important to the
subject of this work, since it contributes 61\% to the static
polarizability  and 98\% to the dynamic correction to the BBR shift
of the $5s5p~^3$P$_0^o$ state. In fact, the accidentally small value
of this transition energy is the source of the anomalously large
(7\%) dynamic correction to the BBR shift of the $^1\!S_0 \rightarrow
\, ^3\!P^o_0$ transition in Sr. We see considerable improvement of
the accuracy in this transition energy from the CI+MBPT to
CI+all-order approximation, by a factor of 6. The CI+MBPT and
CI+all-order values differs from the experiment by  11\% and 1.7\%,
respectively.
\begin{table*}
\caption{\label{table3} Contributions to the  $5s^2\;^1$S$_0$ and $5s5p\;^3$P$_0^o$ static  polarizabilities of Sr in
a.u. The dominant contributions to the valence polarizabilities are listed separately with the corresponding absolute
values of electric-dipole reduced matrix elements given in columns labeled $D$. The theoretical and experimental
\cite{RalKraRea11} transition energies are given in columns $\Delta E_{\rm th}$  and $\Delta E_{\rm expt}$. The
remaining contributions to valence polarizability are given in rows Other. The contributions from the core and
$\alpha_{vc}$ terms are listed together in rows Core + Vc.
 The dominant contributions to $\alpha_0$, listed in columns
$\alpha_0[\mathrm{A}]$ and $\alpha_0(\mathrm{B})$, are calculated with CI + all-order +RPA (no other corrections)
matrix elements and theoretical [A] and experimental [B] energies \cite{RalKraRea11}, respectively. The dominant
contributions to $\alpha_0$ listed in column $\alpha_0[\mathrm{C}]$ are calculated with experimental energies and our
final \textit{ab initio} matrix elements. The dc Stark shift for the $5s5p ~^3$P$_0^o - 5s^2 ~^1$S$_0$ transition is
listed in the last rows of the table.  }
\begin{ruledtabular}
\begin{tabular}{llccrrrrc}
\multicolumn{1}{l}{State} &\multicolumn{1}{l}{Contribution} & \multicolumn{1}{c}{$\Delta E_{\rm th}$} & \multicolumn{1}{c}{$\Delta E_{\rm expt}$} & \multicolumn{1}{c}{$D^{(a)}$} &
\multicolumn{1}{c}{$\alpha_0[\mathrm{A}]$}
&\multicolumn{1}{c}{$\alpha_0[\mathrm{B}]$} &
\multicolumn{1}{c}{$D^{(b)}$}&
\multicolumn{1}{c}{$\alpha_0[\mathrm{C}]$} \\ \hline \\ [-0.3pc]
 $5s^2 ~^1$S$_0$ &$5s^2 ~^1$S$_0 - 5s5p ~^1$P$_1^o$& 21823 &21698  &  5.272  &  186.4  &  187.4  &5.208 & 182.9  \\
                &$5s^2 ~^1$S$_0 - 5s5p ~^3$P$_1^o$& 14739 &14504  &  0.158  &  0.25   &  0.25   &      &   0.25  \\
                &$5s^2 ~^1$S$_0 - 5s6p ~^1$P$_1^o$& 34308 &34098  &  0.281  &  0.34   &  0.34   &      &   0.34  \\
                &$5s^2 ~^1$S$_0 - 4d5p ~^1$P$_1^o$& 41242 &41172  &  0.517 &  0.95    &  0.95   &      &   0.95  \\
&Other  &   &    &       &  4.60    &  4.60  &      &   4.60         \\
&Core + Vc  &    &   &   &  5.29    &  5.29  &      &   5.29         \\
&Total   &  &  &         &197.8     &198.9   &      & 194.4         \\
 &Recomm.$^{(c)}$&  &  &        &           &         &    & 197.14(20) \\
 \hline \\    [-0.3pc]
 $5s5p ~^3$P$_0^o$   &$5s5p ~^3$P$_0 - 5s4d ~^3$D$_1$  & 3777  &  3842  &  2.712  & 285.0 & 280.2 & 2.667 &  270.9 \\
                     &$5s5p ~^3$P$_0^o - 5s6s ~^3$S$_1$& 14673 &  14721 &  1.970  &  38.7 &  38.6 & 1.940 &  37.4 \\
                     &$5s5p ~^3$P$_0^o - 5s5d ~^3$D$_1$& 20660 &  20689 &  2.460  &  42.9 &  42.8 & 2.432 &  41.8 \\
                     &$5s5p ~^3$P$_0^o - 5p^2 ~^3$P$_1$& 21208 &  21083 &  2.619  &  47.3 &  47.6 & 2.620 &  47.6 \\
                     &$5s5p ~^3$P$_0^o - 5s7s ~^3$S$_1$& 23056 &  23107 &  0.516  &   1.69&   1.69&       &   1.69\\
                     &$5s5p ~^3$P$_0^o - 5s6d ~^3$D$_1$& 25326 &  25368 &  1.161  &   7.8 &   7.8 &       &   7.8 \\
                                                  &Other  &      & &         &  29.1 &  29.1 &       &  29.1       \\
                                                  &Core +Vc   &      & &     &   5.55&   5.55&       &   5.55       \\
                                                  &Total   &      & &        & 458.1 & 453.4 &       & 441.9        \\
      &Recomm.$^{(d)}$&  &  &        &           &         &   &   444.51(20) \\
                                                    \hline \\ [-0.3pc]
                                                            $^3$P$_0^o - ^1$S$_0$ &&&&& 260.3 & 254.5 &       & 247.5   \\
Theory~\cite{PorDer06BBR}                       &&&&& 261(4)&       &       &            \\
 Expt.~\cite{MidFalLis12} &&&&&&           &       & 247.374(7)  \\
\end{tabular}
$^{a}$ CI+all-order+RPA values (no other corrections).  $^{b}$ CI+all-order+RPA + other corrections. $^{c}$Obtained
using experimental $5s5p ~^1$P$_1^o$ lifetime from~\cite{YasKisTak06}. $^{d}$Obtained using recommended value for the
$^1$S$_0$ polarizability and the experimental value of the Stark shift~\cite{MidFalLis12}.
\end{ruledtabular}
\end{table*}
\section{\textit{Ab initio} calculation of electric-dipole matrix elements}

 The reduced electric-dipole matrix elements are obtained with the CI+all-order wave functions and
effective  electric-dipole operator $D_{\rm eff}$ in the random-phase
approximation (RPA). The effective operator accounts for the
core-valence correlations in analogy with the effective Hamiltonian
\cite{DzuKozPor98,PorRakKoz99P}.
 We include additional corrections beyond RPA in the calculation of the E1 matrix
 elements
 in comparison with ~\cite{PorDer06BBR,PorLudBoy08}. These contributions include  the
core-Brueckner ($\sigma$), two-particle (2P) corrections, structural radiation (SR), and normalization (Norm)
corrections \cite{DzuKozPor98,PorRakKoz99P}. While we find some cancelation between the various corrections, these
cannot be omitted at the 1\% level of accuracy.  Partial cancelation of the structural radiation and normalization
corrections was discussed in Ref.~\cite{DzuFlaSil87}. Detailed analysis of the structure radiation correction was
carried out in the same work ~\cite{DzuFlaSil87}.

 The results for several transitions that give dominant contributions to the $^1$S$_0 - ^3$P$_0^o$ dc Stark shift are
summarized in Table~\ref{table2}. The percentage differences between the CI+all-order+RPA and CI+MBPT+RPA calculations are given in the column labeled ``Higher orders''.
 We note that it is positive for some transitions
and negative for other transitions. Our final \textit{ab initio} values are given in column labeled ``Final''. We find
that total relative size of corrections beyond CI+all+RPA given in column labeled ``Corr'' is small, 0.04-1.7\%, but
significant. We estimate the uncertainties in the \textit{ab initio} values of the matrix elements to be 1\% based on
the comparison of the CI+MBPT+RPA and CI+all-order+RPA values and combined size of other corrections.

 We also provide the recommended values for these
 transitions. The recommended value for the  $5s^2~^1$S$_0 -
5s5p~^1$P$_1^o$ matrix element was obtained in \cite{PorDer06BBR,PorLudBoy08} from the $^1\!P_1^o$ lifetime measurement
from photoassociation  spectra \cite{YasKisTak06}, the recommended values for all other transitions are obtained in the
present work in Section~\ref{recomm}.

\section{Polarizabilities}
  We evaluated the static and dynamic
polarizabilities of the $5s^2~^1$S$_0$ and $5s5p~^3$P$_0^o$ states of
Sr using the high-precision relativistic CI+all-order method.
 The scalar polarizability $\alpha_0(\omega)$
 is separated into a valence polarizability $\alpha_0^v(\omega)$, ionic core
polarizability $\alpha_c$, and a small term $\alpha_{vc}$
 that modifies ionic core polarizability due to
 the presence of two valence
electrons. The valence part of the polarizability is determined  by solving the inhomogeneous
 equation in valence space, which is approximated  as
\cite{KozPor99a}
\begin{equation}
(E_v - H_{\textrm{eff}})|\Psi(v,M^{\prime})\rangle = D_{\textrm{eff}} |\Psi_0(v,J,M)\rangle
\end{equation}
for the state  $v$ with total angular momentum $J$ and projection $M$. The wave function $\Psi(v,M^{\prime})$ is
composed of parts that have angular momenta of $J^{\prime}=J,J \pm 1$ that allows us to determine the scalar and tensor
polarizability of the state $|v,J,M\rangle$ \cite{KozPor99a}. The effective dipole operator $D_{\textrm{eff}}$ includes
RPA corrections. The core and $\alpha_{vc}$ terms are evaluated in the  random-phase approximation. Their uncertainty
is determined by comparing the DF and RPA values. The small $\alpha_{vc}$ term is calculated by adding $\alpha_{vc}$
contributions from the individual electrons, i.e. $\alpha_{vc}(5s^2)=2\alpha_{vc}(5s)$, and
$\alpha_{vc}(5s5p)=\alpha_{vc}(5s)+\alpha_{vc}(5p)$. The frequency dependence of the core and $\alpha_{vc}$ terms is
negligible, and we use their static values in all calculations.
\begin{table}
\caption{\label{table4} Breakdown of the contributions to the
$5s5p\;^3$P$_0^o$ static polarizability $\alpha_0(\omega=0)$ and
dynamic polarizability  $\alpha_0(\omega)$ at the 813.4~nm magic
wavelength. The dominant contributions to the valence
polarizabilities are obtained with experimental energies and
recommended values of the matrix elements. The electric-dipole
reduced matrix elements are given in column labeled ``$D^{\rm
recom}$''.
 The experimental \cite{RalKraRea11} transition energies are given in column labeled ``$\Delta E_{\rm expt}$''.
  The remaining contributions to valence polarizability are given in row labeled ``Other''.
   The contributions from the core and $\alpha_{vc}$ terms are listed together in row labeled ``Core + Vc''.}
\begin{ruledtabular}
\begin{tabular}{lcccr}
 \multicolumn{1}{l}{Contribution} &  \multicolumn{1}{c}{$\Delta E_{\rm expt}$} & \multicolumn{1}{c}{$D^{\rm recom}$}
 &\multicolumn{1}{c}{$\alpha_0(\omega)$} & \multicolumn{1}{c}{$\alpha_0(\omega=0)$}
\\  \hline
        $5s5p ~^3$P$_0^o - 5s4d ~^3$D$_1$& 3842  &    2.675(13) &     -29.5 &272.6(3.3) \\
        $5s5p ~^3$P$_0^o- 5s6s ~^3$S$_1$& 14721 &     1.962(10) &     126.4 &  38.3(4)   \\
        $5s5p ~^3$P$_0^o - 5s5d ~^3$D$_1$& 20689 &    2.450(24) &     65.6  & 42.5(8)    \\
        $5s5p ~^3$P$_0^o - 5p^2 ~^3$P$_1$& 21083 &    2.605(26) &     71.4  & 47.1(9)    \\
        $5s5p ~^3$P$_0^o - 5s7s ~^3$S$_1$& 23107 &   0.516(8)   &     2.4   &  1.69(5)   \\
        $5s5p ~^3$P$_0^o - 5s6d ~^3$D$_1$& 25368 &   1.161(17)  &     10.2  &  7.8(2)    \\
         Other                       &       &                  &     34.1  &   29.1(9)           \\
        Core + Vc                    &       &                  &    5.55   &   5.55(6)           \\
         Total                       &       &                 &     286.0  &   444.5              \\
\end{tabular}
\end{ruledtabular}
\end{table}

While we do not use the sum-over-states approach in the calculation
of the polarizabilities, it is important  to establish
 the dominant contributions to the final values. We
combine  the electric-dipole matrix elements and energies  according
to the sum-over-states formula for the valence
polarizability~\cite{MitSafCla10}:
\begin{equation}\label{genpol}
\alpha_0^v(\omega) = \frac{2}{3(2J+1)}\sum_n\frac{(E_n-E_v)|\langle v\| D\|
n\rangle|^2}{(E_n-E_v)^2-\omega^2}
\end{equation}
to calculate the contributions of specific transitions.
 Here,  $J$ is the total angular momentum of the state $v$, $D$ is the electric-dipole operator,
  $E_i$ is the energy of the state $i$,
 and frequency $\omega$ is zero in the static polarizability calculations.

\begin{figure}[h]
  \includegraphics[width=2.6in]{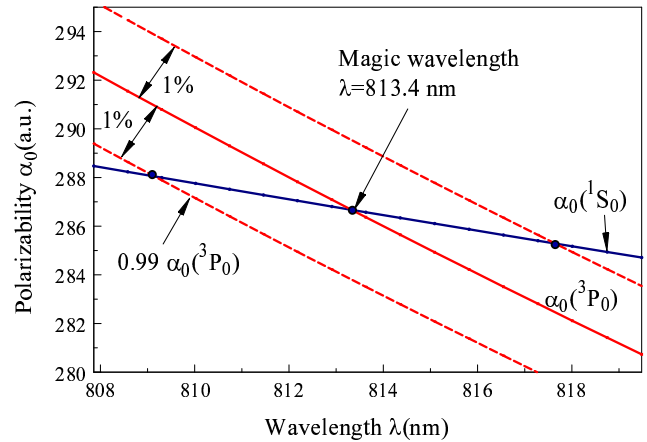}
  \caption{(Color online) The frequency-dependent polarizabilities of the Sr $5s^2
   ~^1$S$_0$ and $5s5p ~^3$P$_0^o$ states near
  813.4~nm magic wavelength. The frequency-dependent polarizability of $5s5p ~^3$P$_0^o$ shifted by $\pm$ 1\%
  are shown  to illustrate the sensitivity of the magic wavelength  to the $5s5p ~^3$P$_0^o$ polarizability.
     The magic wavelength is marked with arrow.  }
  \label{fig1}
\end{figure}

\begin{table*} [ht]
\caption{\label{table5} Dynamic corrections to the BBR shift of the
$5s^2 ~^1$S$_0- 5s5p\;^3$P$^o_0$ clock transition in Sr at $T=300~K$
(in Hz). Quantities $\eta_i$, $\Delta E_{g}^{\rm dyn}$, and $\Delta
\nu_{^3\!P_0^o - ^1\!S_0}^{\rm dyn}$ are defined in text. }
\begin{ruledtabular}
\begin{tabular}{lrrrrcc}
 \multicolumn{1}{l}{}&
\multicolumn{1}{c}{$\eta_1$} & \multicolumn{1}{c}{$\eta_2$}&
\multicolumn{1}{c}{$\eta_3$} & \multicolumn{1}{c}{$\eta$} &
 \multicolumn{1}{c}{$\alpha_0(\omega=0)$} & \multicolumn{1}{c}{$\Delta E_{g}^{\rm dyn}/h$}  \\  \hline     \\[-0.3pc]
  Total $(5s^2 ~^1$S$_0)$    &   0.00163 &0.00001 &0&0.00164&   197.1&  -0.0028              \\    [0.5pc]

   $5s5p ~^3$P$_0^o - 5s4d ~^3$D$_1$& 0.03394 &0.00414 & 0.00088    & 0.03896&         &         \\
   $5s5p ~^3$P$_0^o - 5s6s ~^3$S$_1$& 0.00032 &0       & 0          & 0.00032&         &         \\
   $5s5p ~^3$P$_0^o - 5s5d ~^3$D$_1$& 0.00018 &0       & 0          & 0.00018&         &         \\
   $5s5p ~^3$P$_0^o - 5p^2 ~^3$P$_1$& 0.00019 &0       & 0          & 0.00020&         &         \\
   $5s5p ~^3$P$_0^o - 5s7s ~^3$S$_1$& 0.00001 &0       & 0          & 0.00001&         &         \\
   $5s5p ~^3$P$_0^o - 5s6d ~^3$D$_1$& 0.00002 &0       & 0          & 0.00002&         &          \\
     Total $(5s5p ~^3$P$_0^o)$        & 0.03467 &0.00415 & 0.00088    & 0.03970&  444.6 &   -0.1520    \\  [0.5pc]\hline
     \\[-0.3pc]
    Final $\Delta \nu_{^3\!P_0^o - ^1\!S_0}^{\rm dyn}$
                                  &            &       &         &         &  &  -0.1492(16) \\
    Ref.~\cite{MidFalLis12}       &            &       &         &         &  & -0.1477(23)  \\
\end{tabular}
\end{ruledtabular}
\end{table*}

We have carried out several calculations of the dominant contributions to the polarizabilities using different sets of
the energies and E1 matrix elements in order to understand the difference of the theoretical predictions for the Stark
shift $5s^2~^1\!S_0 - 5s5p~^3\!P^o_0$ $\Delta \alpha = 261(4)~$a.u. and recent experimental measurement $\Delta \alpha
= 247.374(7)$~a.u. as well as to provide a recommended value for the $5s5p~ ^3$P$_0^o- 5s4d~ ^3$D$_1$ matrix element.
The results are summarized in Table~\ref{table3}. Other theoretical calculations of Sr polarizabilities were recently
compiled in review \cite{MitSafCla10}.
  The ground-state polarizability of Sr was calculated using relativistic coupled-cluster (RCC)
  method in \cite{SahDas08}. Their value 199.7(7.3)~a.u. is in agreement with our calculations.

In Table~\ref{table3} the absolute values of the corresponding reduced electric-dipole matrix elements are listed in
columns labeled ``$D$'' in a.u.. The theoretical and experimental \cite{RalKraRea11} transition energies  are given in
columns $\Delta E_{\rm th}$ and $\Delta E_{\rm expt}$. The remaining valence contributions are given in rows labeled
``Other''.
 The contributions from the core and $\alpha_{vc}$ terms are
listed together in row labeled ``Core +Vc''. The dominant contributions to $\alpha_0$ listed in columns
$\alpha_0[\mathrm{A}]$ and $\alpha_0[\mathrm{B}]$ are calculated with CI + all-order +RPA (no other corrections) matrix
elements and theoretical [A] and experimental [B] energies \cite{RalKraRea11}, respectively.

Our $\alpha_0[\mathrm{A}]$  result agrees with the earlier calculation of ~\cite{PorDer06BBR} which was carried out
using CI+MBPT approach with energy fitting that approximated missing higher-order corrections to the wave functions.
 We note that this may be fortuitous since the calculation of ~\cite{PorDer06BBR} was carried out in $V^N$ potential,
 while we are using $V^{N-2}$ potential since the present version of the CI+all-order method is formulated for $V^{N-2}$
potential.  The E1 matrix elements in \cite{PorDer06BBR,PorLudBoy08} included RPA but omitted all other corrections
calculated in the present work. We find that replacing the theoretical energies with experimental values reduces the
Stark shift by 2.3\%. We note that in the case of Sr all of the states contributing to the polarizabilities are
included in our computational basis and this procedure is not expected to cause problems with basis set completeness,
as in the case of Yb ~\cite{DzuDer10}.
 The dominant contributions to $\alpha_0$ listed in column
$\alpha_0[\mathrm{C}]$  are calculated with experimental energies and final ab initio  matrix elements. Inclusion of
the small corrections further reduces the value of the Stark shift by 3.1\%, and our resulting value obtained with our
final \textit{ab initio} matrix elements is in excellent agreement with experiment \cite{MidFalLis12}.

        \section{Determination of recommended values of electric-dipole matrix elements}
\label{recomm}
 To further improve our values of the other matrix elements, we
use the measurement of the  magic wavelength for  the $^3$P$_0^o -
^1$S$_0$ clock transition to determine recommended values of the
$5s5p ~^3$P$_0^o - 5s6s ~^3$S$_1$, $5s5p ~^3$P$_0^o - 5s5d ~^3$D$_1$,
and $5s5p ~^3$P$_0^o - 5p^2 ~^3$P$_1$ matrix elements. The $^1$S$_0$
polarizability at the 813.4~nm magic wavelength is 286.0~a.u. which
is essentially fixed by the value of the $5s5p ~^1$P$_1^o$ lifetime
~\cite{PorLudBoy08}. The contributions to the $^3$P$_0^o$
polarizability at the magic wavelength are listed in
Table~\ref{table4}. Since the contribution of the $5s5p ~^3$P$_0^o-
5s6s ~^3$S$_1$  transition is dominant, the magic wavelength limits
the value of this matrix element within about 0.5\%. Since we appear
to systematically overestimate the correction to matrix elements
beyond CI+all+RPA approximation, we adjust the values of two other
$5s5p ~^3$P$_0^o - 5s5d ~^3$D$_1$ and $5s5p ~^3$P$_0^o - 5p^2
~^3$P$_1$ matrix elements in a similar way as the $5s5p ~^3$P$_0^o-
5s6s ~^3$S$_1$ one.
 We plot the dynamic polarizabilities of the $^1$S$_0$ and $^3$P$_0^o$
 states in the vicinity
of the magic wavelength on Fig.~\ref{fig1} to illustrate that the
crossing point is extremely sensitive to the matrix
element values.
 The 0.5\% change in the
values of the matrix elements (corresponding to 1\% change in the value of the $^3\!P_0^o$ polarizability)  shifts the
crossing point by more than 4~nm.

\begin{table*}
\caption{ \label{table6} Experimental  transition energies (in
cm$^{-1}$), theoretical line strengths (in a.u.), transition rates
(in s$^{-1}$), and branching ratios for transitions contributing to
the $5s4d\ ^3$D$_J$ lifetimes. The CI+RPA, CI+MBPT+RPA, and
CI+all-order+RPA results are listed in columns labeled ``CI'',
``MBPT'', and ``All'', respectively. The recommended values of the
$\langle 5s5p\ ^3$P$^o_{1,2} ||D|| 5s4d\ ^3$D$_{1,2} \rangle|$
 and $|\langle 5s5p\ ^3$P$^o_{2} ||D|| 5s4d\ ^3$D$_{3} \rangle|$
matrix elements are obtained using the recommended matrix element for
the $5s4d\ ^3$D$_1 \rightarrow 5s5p\ ^3$P$^o_0$ transition (i.e.
scaled by 0.9862). Experimental energies are used in all cases.
Numbers in square brackets represent powers of 10.}
\begin{ruledtabular}
\begin{tabular}{lccccccccccc}
\multicolumn{1}{c}{Transition}& \multicolumn{1}{c}{$\Delta E_{\rm
epxt}$}& \multicolumn{4}{c}{Line strengths $S$}&
\multicolumn{3}{c}{Transition rates $A_{ab}$}&\multicolumn{3}{c}{Branching ratios}\\
\multicolumn{2}{c}{}&
 \multicolumn{1}{c}{CI}&\multicolumn{1}{c}{MBPT}& \multicolumn{1}{c}{All}&   \multicolumn{1}{c}{Recomm.}
&\multicolumn{1}{c}{MBPT}& \multicolumn{1}{c}{All}&
\multicolumn{1}{c}{Recomm.}&
   \multicolumn{1}{c}{CI}&\multicolumn{1}{c}{MBPT}& \multicolumn{1}{c}{All}\\
\hline
$^3$D$_1 \rightarrow \, ^3$P$_0^o$&    3842& 9.503& 7.189& 7.357&  7.156&   2.753[5]&  2.817[5]& 2.740[5]&0.5949 &  0.5954 &     0.5953  \\
$^3$D$_1 \rightarrow \, ^3$P$_1^o$&    3655& 7.172& 5.414& 5.543&  5.391&   1.785[5]&  1.828[5]& 1.777[5]&0.3866 &  0.3861 &     0.3862  \\
$^3$D$_1 \rightarrow \, ^3$P$_2^o$&    3260& 0.485& 0.365& 0.374&  0.364&   8.541[3]&  8.750[3]& 8.510[3]&0.0186 &  0.0185 &     0.0185  \\
$\sum_{b\leq a}{A_{ab}}$    &        &      &      &      &       &
4.623[5]&  4.722[5]&       4.602[5]&      &          &
\\ [-0.6pc]
 \\
  $^3$D$_2 \rightarrow \, ^3$P$_1^o$&    3714& & &16.605& 16.149 &  &  3.448[5]& 3.354[5]& &   &   0.8058  \\
  $^3$D$_2 \rightarrow \, ^3$P$_2^o$&    3320& & &5.602 & 5.449 &   &  0.831[5]& 0.808[5]& &   &   0.1942  \\
  $\sum_{b\leq a}{A_{ab}}$    &        &      &      &      &       && 4.279[5]& 4.162[5]&      &          &
  \\ [-0.6pc]\\
 $^3$D$_3 \rightarrow \, ^3$P$_2^o$&  3421  & & & 31.519& 30.655 &  &  3.652[5]& 3.552[5]& &   &
 \\
\end{tabular}
\end{ruledtabular}
\end{table*}

\begin{table}
\caption{ \label{table61}  The lifetimes of the $5s4d ~^3$D$_J$ states
 in ns. The last three rows give term-averaged $^3$D lifetime.}
\begin{ruledtabular}
\begin{tabular}{lccc}   &
\multicolumn{1}{c}{MBPT}& \multicolumn{1}{c}{All}&
\multicolumn{1}{c}{Recomm.}\\
\hline
  $\tau(5s4d~ ^3$D$_1)$  & 2163 &      2113 & 2171(24)   \\
   $\tau(5s4d~ ^3$D$_2)$  &     &      2337& 2403(27)    \\
 $\tau(5s4d~ ^3$D$_3)$    &     &      2738 & 2816(31)
 \\
 \\[-0.6pc]  \hline  \\
 $\tau(5s4d~ ^3$D$)$       &     &      2453& 2522(28)
 \\
 Expt.~\cite{MilYuCoo92}   &    & 2900(200) &     \\
  Expt.~\cite{RedSanEci04}  &   & 2500(200) &
 \\
\end{tabular}
\end{ruledtabular}
\end{table}

 After we determined the values of these three matrix elements, we used them to
obtain the value of the $5s5p ~^3$P$_0^o - 5s4d~ ^3$D$_1$ matrix element from the experimental value of the Stark
shift. We determine its uncertainty from the uncertainty of all the other contributions to the $^3$P$_0^o$
polarizability value (listed in the last column of Table~\ref{table4}). Since our theoretical values may experience a
systematic shift in one direction, we (somewhat conservatively) simply add all of the uncertainties, totaling to
$3.3$~a.u, instead of adding them in quadrature. Assigning  this value to be the uncertainty in the dominant  $^3\!D_1$
contribution of 272.7~a.u., we estimate the uncertainty in the recommended value of the $5s5p~^3$P$_0^o - 5s4d~^3$D$_1$
matrix element to be 0.5\%. Since the contributions to both static and dynamic polarizabilities from $5s5p ~^3$P$_0^o -
5s7s ~^3$S$_1$ and $5s5p ~^3$P$_0^o - 5s6d ~^3$D$_1$ transitions are small, we use \textit{ab initio} CI+all+RPA values
and assign them 1.5\% uncertainty. Combined with experimental energies and other small contributions, the set of
recommended matrix elements reproduces  recommended values for both the $5s5p~^3$P$_0^o$ static and dynamic
polarizability at 813.4~nm magic wavelength.

\section {Blackbody radiation shift}
The leading contribution to the multipolar black body radiation (BBR)
shift of the energy level $g$ can be expressed in terms of the
electric dipole transition matrix elements~\cite{FarWin81}
\begin{eqnarray}
\Delta E_{g}= -\frac{(\alpha k_B T)^3}{2J_{g}+1} \sum_n |\langle g||D||n \rangle|^2 F_1(y_n). \label{dEg_F1}
\end{eqnarray}
Here $k_B$ is the Boltzmann constant, $y_n \equiv (E_n - E_g)/(k_B T)$, and $F_1(y)$ is the function introduced by
Farley and Wing in~\cite{FarWin81}. Its asymptotic expansion is given by
\begin{eqnarray}
F_{1}\left( y\right) \approx \frac{4\pi ^{3}}{45y}+%
\frac{32\pi ^{5}}{189y^{3}}+ \frac{32\pi ^{7}}{45y^{5}}+ \frac{512\pi ^{9}}{99y^{7}}.
\end{eqnarray}

The \eref{dEg_F1} can be expressed in terms of the dc polarizability
 $\alpha_g(\omega=0)$ of the state $g$
as~\cite{PorDer06BBR}
\begin{equation}
\Delta E_{g} = -\frac{2}{15} (\alpha \pi)^3 (k_B T)^4 \alpha_g(0) + \Delta E_{g}^{\rm dyn}. \label{delEg}
\end{equation}
Here $\Delta E_{g}^{\rm dyn}$ is determined as
\begin{equation}
\Delta E^{\rm dyn}_{g} \equiv -\frac{2}{15} (\alpha \pi)^3 (k_B T)^4 \alpha_g(0) \, \eta \label{delEg_dyn}
\end{equation}
and $\eta$ represents a ``dynamic'' fractional correction to the total shift that reflects the averaging of the
frequency dependence of the polarizability over the frequency of the blackbody radiation spectrum. Corresponding shift
in the clock transition frequency, $\Delta \nu_{^3\!{\rm P}_0^o - ^1\!S_0}^{\rm dyn}=(\Delta E^{\rm
dyn}_{^3\!P_0^o}-\Delta E^{\rm dyn}_{^1\!S_0})/h$, is referred to as dynamic correction to the BBR shift.
 The quantity $\eta$ can be approximated
by~\cite{PorDer06BBR}
\begin{eqnarray}
&&\eta= \eta_1+\eta_2+\eta_3=\frac{80}{63\,(2J_g+1)} \,
\frac{\pi^2}{\alpha_g(0) k_B T} \nonumber \\
&&\times
  \sum_n \frac{ |\langle n||D||g \rangle|^2 }{y_n^3} \left(
1+ \frac{21 \pi^2}{5\,y_n^2} + \frac{336 \pi^4}{11y_n^4} \right) . \label{eta}
\end{eqnarray}

Dynamic corrections to the BBR shift of the $5s^2 ~^1$S$_0- 5s5p\;^3$P$_0^o$ clock transition in Sr at $T=300~K$ (in
Hz) are given in Table~\ref{table5}. The dynamic correction to the BBR shift of the $^3$P$_0^o$ level is dominated by
the contribution from the
 $5s5p~ ^3$P$_0^o  - 5s4d ~^3$D$_1$ transition, which contributes 98.2\%
 of the total. Our final
result $-0.1492(16)$~Hz is in excellent agreement with recent value $-0.1477(23)$~Hz of Ref.~\cite{MidFalLis12}.

Our result enables us to propose an approach for further reduction of the
uncertainty in the BBR shift: a measurement
of the $5s4d~^3$D$_1$ lifetime with 0.5\% uncertainty would provide
 the value of the BBR shift in Sr clock that is
accurate to about 0.5\%, which would be  a factor of 2 improvement
in the uncertainty stated here. Such a determination
assumes accurate knowledge of the branching ratios.

The $5s4d ~^3$D$_1$ level decays to all three $5s5p~^3$P$_{0,1,2}^o$ states,
but the branching ratio to the $^3$P$_2^o$
level is very small. The lifetime of a state $a$ is calculated
as $$ \tau_a=\frac{1}{\sum_{b\leq a}A_{ab}}.
$$
The E1 transition rates $A_{ab}$ are calculated using
$$
A_{ab}=\frac{2.02613\times 10^{18}}{\lambda^3}\frac{S}{2J_a+1} {\rm s}^{-1},
$$
where $\lambda$ is the wavelength of the transition in $\AA$~and $S$ is the line strength.

We find that the branching ratios are essentially independent of the
correlation corrections to the matrix elements.  We note that line
strength ratios are close to the non-relativistic ones (5/9, 5/12,
1/36), with the differences being $-$0.23\%, +0.22\%, and 1.4\% for
the $^3$D$_1-^3$P$_J$ transitions, respectively. We illustrate this
point in Table~\ref{table6}, where we list the relevant energies,
line strengths $S$,  transition rates $A$, and branching ratios  in
the CI, CI+MBPT, and CI+all-order approximations.

The RPA corrections to the matrix elements are included in all cases.
We used experimental energies in all calculations for consistency.
 We find that the difference in the CI, CI+MBPT, and CI+all-order
branching ratio results is less than 0.1\%. Since all the other
corrections are small, their uncertainties should be even smaller. As
a result, the accuracy of our branching ratios should be better than
0.2\%. The recommended values for the $5s5p\ ^3$P$_{1,2}^o - 5s4d\
^3$D$_1$ matrix elements are obtained using the recommended matrix
element for the $5s5p\ ^3$P$_0^o - 5s4d\ ^3$D$_1$ transition and
CI+all-order branching ratios. The recommended values for the
transition rates and the $5s4d\ ^3$D$_1$ lifetime, 2172(24)~ns, are
obtained using the
  recommended values of the matrix elements and experimental energies.
We also list the recommended values $^3$D$_2$, $^3$D$_3$, and
term-averaged $^3$D lifetimes in Table~\ref{table61}. The $^3$D
term-averaged lifetime is compared with
experiment~\cite{MilYuCoo92,RedSanEci04}.

\section{Conclusion}

We have evaluated the static and dynamic polarizabilities of the $5s^2~^1$S$_0$ and $5s5p~^3$P$_0^o$ states of Sr and
explained the discrepancy between the recent experimental $5s^2~^1$S$_0 - 5s5p~^3$P$_0^o$ dc Stark shift measurement
~\cite{MidFalLis12} and the earlier theoretical result \cite{PorDer06BBR}. Our theoretical value for the dc Stark shift
of the clock transition, 247.5~a.u., is in excellent agreement with the experimental result. We have provided the
recommended values of the matrix elements for transitions that give dominant contributions to the clock Stark shift and
evaluated their uncertainties.  We evaluated the dynamic correction to the BBR shift of the $^1$S$_0 -\, ^3$P$_0^o$
clock transition at $300~K$ to be -0.1492(16)~Hz and proposed an approach for further reduction of the uncertainty in
the BBR shift.

\section*{Acknowledgements}

This research was performed under the sponsorship of the U.S. Department of Commerce, National Institute of
 Standards and Technology, and was supported by the National Science Foundation under Physics Frontiers Center Grant
 No. PHY-0822671 and by the Office of Naval Research. The work of S.G.P. was supported in part by US NSF
Grant No.\ PHY-1068699 and RFBR Grant No. 11-02-00943.
The work of M.G.K was supported in part by RFBR Grant No. 11-02-00943.


\end{document}